\documentclass[manuscript]{emulateapj}
\usepackage{natbib}
\slugcomment{Draft version \today}

\shorttitle{{\it Chandra} observations of SN 1987A}
\shortauthors{Helder et al.}

\begin{document}

\title{Chandra observations of SN 1987A: the soft X-ray light curve revisited}

\author{E.A. Helder$^1$, P.S. Broos$^1$, D. Dewey$^2$, E. Dwek$^3$, R. McCray$^4$, \\S. Park$^5$, J.L. Racusin$^6$, S.A. Zhekov$^7$, D.N. Burrows$^1$}
\affil{$^1$ Department of Astronomy and Astrophysics, The Pennsylvania State University, 525 Davey Lab, University Park, PA 16802 USA}
\affil{$^2$ MIT Kavli Institute, Cambridge, MA 02139, USA}
\affil{$^3$ Observational Cosmology Lab, Code 665, NASA Goddard Space Flight Center, Greenbelt, MD 20771, USA}
\affil{$^4$ JILA, University of Colorado and NIST, 440 UCB, Boulder, CO 80309, USA}
\affil{$^5$ Department of Physics, University of Texas at Arlington, Box 19059, Arlington, TX 76019, USA}
\affil{$^6$ NASA, Goddard Space Flight Center, Code 661, Greenbelt, MD, USA}
\affil{$^7$ Space Research and Technology Institute, Sofia 1113, Akad. G.Bonchev
str., bl.1, Bulgaria}

\begin{abstract}
We report on the present stage of SN 1987A as observed by the {\it Chandra} X-ray Observatory. We reanalyze published {\it Chandra} observations and add three more epochs of {\it Chandra} data to get a consistent picture of the evolution of the X-ray fluxes in several energy bands. We discuss the implications of several calibration issues for {\it Chandra} data. Using the most recent {\it Chandra} calibration files, we find that the 0.5-2.0 keV band fluxes of SN 1987A have increased by $\sim$ 6 $\times 10 ^{-13}$ erg s$^{-1}$ cm$^{-2}$ per year since 2009. This is in contrast with our previous result that the 0.5-2.0 keV light curve showed a sudden flattening in 2009. Based on our new analysis, we conclude that the forward shock is still in full interaction with the equatorial ring.
\end{abstract}


\section{Introduction}

Stellar winds of massive stars leave their imprints on the surrounding medium before they explode. After the explosion, the outer shocks of the supernova  probe the circumstellar medium, thereby mapping the late stages of the stars evolution. 

Because of its youth, the remnant of SN 1987A provides a unique opportunity to investigate these late stages. The case of SN 1987A is especially interesting, as it exhibits the enigmatic three ring structure visible in optical images \citep{Crotts89}. This structure probably originates from the interaction of a wind from the slow red supergiant phase with the faster wind from the blue supergiant phase \citep{Luo91,Wang92}. 
The morphology of this structure might provide clues to a binary-merger scenario for the progenitor \citep{Blondin93,Morris07,Morris2009}. However, this three ring structure might also result from mass-loss from a fast-rotating star \citep{Chita2008}. Currently, the explosion is sweeping up the inner equatorial ring that was formed by the late stages of the star's evolution.

As the inner equatorial ring is about a hundred times denser than its surroundings \citep[e.g., ][]{Matilla10}, the interaction of the explosion with this structure has a direct impact on the evolution and observational properties of the remnant. In 1995, the outer ejecta began to light up this structure. 
By October 2000, 12 spots had lit up at visible wavelengths \citep{Sugerman2002}. Infrared observations indicate that the full interaction with the ring started around mid-2002 \citep{Dwek2010}. A year and a half later, the expansion in X-rays slowed down dramatically \citep{Racusin2009} while, at the same time, the soft X-ray light curve turned up \citep{Park05}.
 
Based on past {\it Chandra} observations, \cite{Zhekov05} established that the X-rays are emitted from a flat spatial structure. A model consisting of three components has been proposed for interpreting the high resolution X-ray spectroscopy from SN 1987A \citep{Zhekov10,Dewey12}. The first component is emitted by material in the dense ring that has been heated by a slow shock (500 km s$^{-1}$). This component has a low electron temperature ($\sim 0.3 - 0.5$ keV). The second component is emitted by the shock-heated circumstellar medium located above and below the equatorial ring. These shocks have not been slowed down as much and therefore the plasma is hotter. Also, the line emission of this component has to be broad (thousands of km s$^{-1}$). This allowed \cite{Dewey12} to estimate the current contribution of this component to be of the order of $\sim 20$\% of the total emission in the 0.5-2.0 keV band, based on high resolution High Energy Transmission Grating (HETG) spectra and Reflection Grating Spectrometer (RGS) spectra. Further investigation of the X-ray spectra of SN 1987A shows that there is need for a third component with a higher ($\sim$ 3 keV) temperature, but without broad line emission. \cite{Zhekov2006} proposed that this is shocked plasma, reheated by a reflected shock caused by the interaction with the equatorial ring. Alternatively, \cite{Dewey12} proposed that this high temperature component arises from a fast shock, traveling through the uniform equatorial ring material (in contrast to the low temperature component, which they consider being emitted by shocked denser clumps in the equatorial ring). 

The {\it Chandra} X-ray Observatory has observed SN 1987A at a regular pace of about two observations per year, resulting in a denser coverage of the evolution of the remnant than with any other X-ray observatory. \cite{Park2011} summarized the observations through 2010. In the 0.5-2.0 keV light curve, they found a break (i.e., the pace with which the flux increases suddenly slowed down) around 8200 days after the explosion ($\sim$ Y2009). Based on this result, they concluded that the shocks responsible for the soft X-ray emission must have recently encountered less dense material.

However, the May 2012 revision to the ACIS calibration (CALDB 4.4.10)\footnote{\url{http://cxc.harvard.edu/caldb/downloads/Release\_notes/\\ CALDB\_v4.4.10.html}} introduced a significant change to the time-dependent model for the transmission of the optical blocking filter.  The new model predicts lower transmission than the previous model (a 15\% change in the optical depth in 2012), reflecting an apparent increase in the rate of deposition on the filter.

In this paper, we reconsider the light curve break result \citep{Park2011} by reanalyzing previously published epochs using the newest calibration and by adding three recent observations (2011 March and September, 2012 April). We show that the break seen by \cite{Park2011} is due primarily to the (unmodeled) increased rate of deposition on the optical blocking filter and not to a different phase in the evolution of the remnant.

\section{Data and results}

Table \ref{tabel} lists the 23 monitoring observations used for this study. We chose not to use the observations earlier in the mission (ObsIds 1387 and 122 at 4608 and 4711 days after the explosion), as these observations were taken at a focal plane temperature of -110$^{\rm \circ}$ C and no charge transfer inefficiency adjustments are available for observations taken at this temperature\footnote{\url{http://cxc.harvard.edu/ciao4.4/why/cti.html}}. However, we report the spatial distribution of the X-ray data at these epochs in section \ref{imaging}. Since the remnant has thus far brightened by a factor of $\sim$ 34 during the {\it Chandra} mission, the observing configuration was changed several times to mitigate photon pileup effects\footnote{\url{http://cxc.harvard.edu/ciao/ahelp/acis\_pileup.html}}.
 First, this was done by reading out an increasingly smaller subarray of the ACIS-S3 chip. All observations from day 7799 onward used the ACIS-HETG configuration\footnote{
Use of the HETG is an excellent pileup mitigation strategy, since the count rate in the zeroth order is reduced by a factor of 9.
}, which produces both a zeroth-order image and a dispersed spectrum.  The dispersed spectrum provides an invaluable cross-check for our ACIS results.  From day 8434, the observation was moved to the bottom of the chip, to decrease the frame time to 1s. The time of this transition coincides with the accelerated pace of the increase of the contamination on the filter. For the data reduction, we largely follow the methods that were used before on {\it Chandra} data of SN 1987A \citep{Burrows,Racusin2009,Park2011}. We filtered for background flares, and ignored the flags set by the {\it Chandra} afterglow filter (status=000000000000xxxx0000000000000000) as this filter tends to filter real events for bright sources \cite[e.g., ][]{Townsley03}.

\begin{center}
\begin{table*}
\small{
\caption{ACIS-S3 Observations\label{tabel}}
\begin{tabular}{llllll}
ObsId(s) & SN age & Instrument &frame time& 0.5-2.0 keV flux$^a$ & 3.0-8.0 keV flux$^a$\\
 &days & & [seconds] & [$10^{-13}$ erg s$^{-1}$ cm$^{-2}$] & [$10^{-13}$ erg s$^{-1}$ cm$^{-2}$] \\
\hline \\
1967&5036&ACIS-S3&3.2 &$ 2.79_{- 0.13}^{+ 0.11}$ (1.18) & $ 0.67_{- 0.05}^{+ 0.04}$ (0.85) \\
1044&5175&ACIS-S3&3.2 &$ 3.09_{- 0.33}^{+ 0.22}$ (1.17) & $ 1.17_{- 0.34}^{+ 0.09}$ (1.01) \\
2831&5406&ACIS-S3&3.1&$ 4.26_{- 0.27}^{+ 0.27}$ (1.24) & $ 0.91_{- 0.18}^{+ 0.06}$ (0.86) \\
2832&5560&ACIS-S3&3.1&$ 5.26_{- 0.35}^{+ 0.28}$ (1.29) & $ 0.92_{- 0.06}^{+ 0.20}$ (0.71) \\
3829&5790&ACIS-S3&3.1&$ 7.44_{- 0.97}^{+ 0.51}$ (1.36) & $ 1.20_{- 0.60}^{+ 0.02}$ (0.76) \\
3830&5979&ACIS-S3&3.1&$ 9.43_{- 1.01}^{+ 0.58}$ (1.52) & $ 1.46_{- 0.42}^{+ 0.11}$ (0.82) \\
4614&6157&ACIS-S3&3.1&$ 11.59_{- 0.66}^{+ 0.44}$ (1.52) & $ 1.80_{- 0.28}^{+ 0.12}$ (0.83) \\
4615&6359&ACIS-S3&1.5&$ 14.75_{- 0.98}^{+ 0.63}$ (1.29) & $ 1.98_{- 0.32}^{+ 0.14}$ (0.88) \\
5579 \& 6178&6530&ACIS-S3&0.4 &$ 17.71_{- 0.65}^{+ 0.61}$ (1.08) & $ 1.98_{- 0.12}^{+ 0.08}$ (0.83) \\
5580 \& 6345&6713&ACIS-S3&0.4 &$ 21.82_{- 0.85}^{+ 0.76}$ (1.11) & $ 2.53_{- 0.25}^{+ 0.11}$ (0.94) \\
6668&6914&ACIS-S3&0.4 &$ 26.72_{- 0.99}^{+ 0.76}$ (1.13) & $ 3.23_{- 0.26}^{+ 0.18}$ (0.96) \\
6669&7094&ACIS-S3&0.4 &$ 31.61_{- 0.78}^{+ 1.08}$ (1.14) & $ 3.11_{- 0.12}^{+ 0.10}$ (0.85) \\
7636&7270&ACIS-S3&0.4 &$ 37.68_{- 0.43}^{+ 0.40}$ (1.18) & $ 3.73_{- 0.17}^{+ 0.10}$ (0.89) \\
7637&7445&ACIS-S3&0.4 &$ 42.27_{- 0.29}^{+ 0.88}$ (1.19) & $ 3.39_{- 0.14}^{+ 0.14}$ (0.77) \\
9142 \& 9806&7625&ACIS-S3&0.2 &$ 46.37_{- 0.46}^{+ 2.15}$ (1.09) & $ 3.42_{- 0.36}^{+ 0.21}$ (0.63) \\
9144&7799&HETG&1.1 &$ 50.36_{- 1.70}^{+ 1.56}$ (1.06) & $ 5.00_{- 0.42}^{+ 0.26}$ (0.92) \\
10221 \& 10852-10855&8000&HETG&1.1 &$ 55.32_{- 11.84}^{+ 2.21}$ (1.09) & $ 4.86_{- 4.06}^{+ 0.13}$ (0.81) \\
10222 \& 10926&8233&HETG&1.1&$ 60.78_{- 3.66}^{+ 2.45}$ (1.07) & $ 5.83_{- 0.72}^{+ 0.39}$ (0.89) \\
12125 \& 12126 \& 11090&8434&HETG&1.0 &$ 63.50_{- 3.48}^{+ 2.62}$ (1.08) & $ 5.95_{- 0.41}^{+ 0.24}$ (0.90) \\
13131 \& 11091&8618&HETG&1.0 &$ 66.15_{- 1.52}^{+ 1.81}$ (1.08) & $ 6.08_{- 0.50}^{+ 0.23}$ (0.85) \\
12539&8796&HETG&1.0&$ 68.59_{- 1.53}^{+ 2.00}$ (1.08) & $ 7.62_{- 0.35}^{+ 0.20}$ (0.98) \\
14344 \& 12540&8976&HETG&1.0&$ 71.08_{- 2.43}^{+ 1.96}$ (1.08) & $ 7.66_{- 0.47}^{+ 0.34}$ (0.88) \\
13735 \& 14417&9175&HETG&1.0 &$ 75.97_{- 1.34}^{+ 1.38}$ (1.11) & $ 8.91_{- 0.65}^{+ 0.24}$ (0.97) \\
\end{tabular}\\
\vskip 0.2mm
Observation log of the {\it Chandra} S3 observations of SN 1987A.  $^a$ Numbers between  parentheses are the ratios of the pileup corrected fluxes and the non-pileup corrected fluxes.
}
\end{table*}
\end{center}

\subsection{Imaging}
\label{imaging}
\begin{figure*}
\epsscale{1.}
\plotone{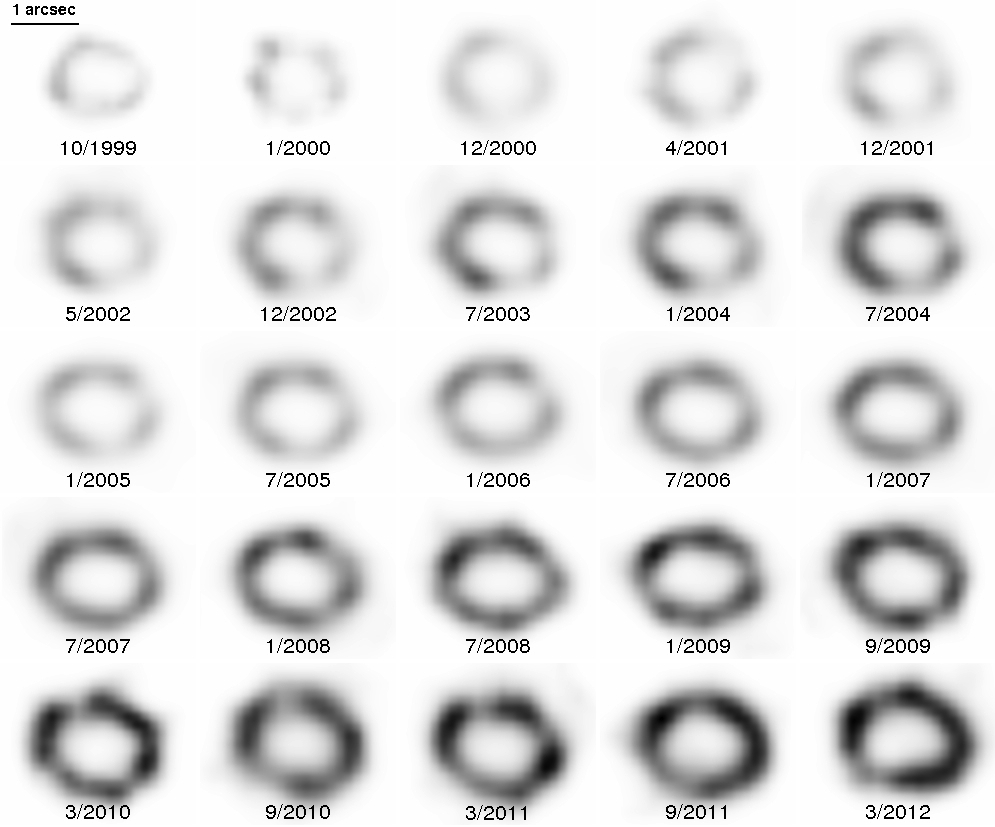}
\caption{Estimated intrinsic morphology (corrected for the {\it Chandra} PSF) of SN 1987A at all epochs utilized in this study, proceeding from upper left to lower right. The image scaling is linear and the brightness scales with the 0.5-8.0 flux in the observation. The scaling limits of the two top rows are offset by a factor of 3 for visibility. In the images, north is up, east is left.}
\label{overview}
\end{figure*}

With an angular diameter of $1\farcs5$, SN 1987A is barely resolved in raw {\it Chandra} ACIS images. To extract as much spatial information from the observations as possible, we adopt
the same approach as in previous studies \citep{Burrows,Racusin2009,Park2011}. We remove the pixel randomization that is usually added by the {\it Chandra} software. Additionally, we improved the spatial resolution using split-pixel events \citep{Mori2001}. We then deconvolve the image for the point spread function of the telescope\footnote{Using MARX version 4.4.0 (space.mit.edu/cxc/marx)} using the Lucy-Richardson iterative deconvolution algorithm \citep{Lucy,Richardson}. 

\begin{figure}
\epsscale{1.}
\plotone{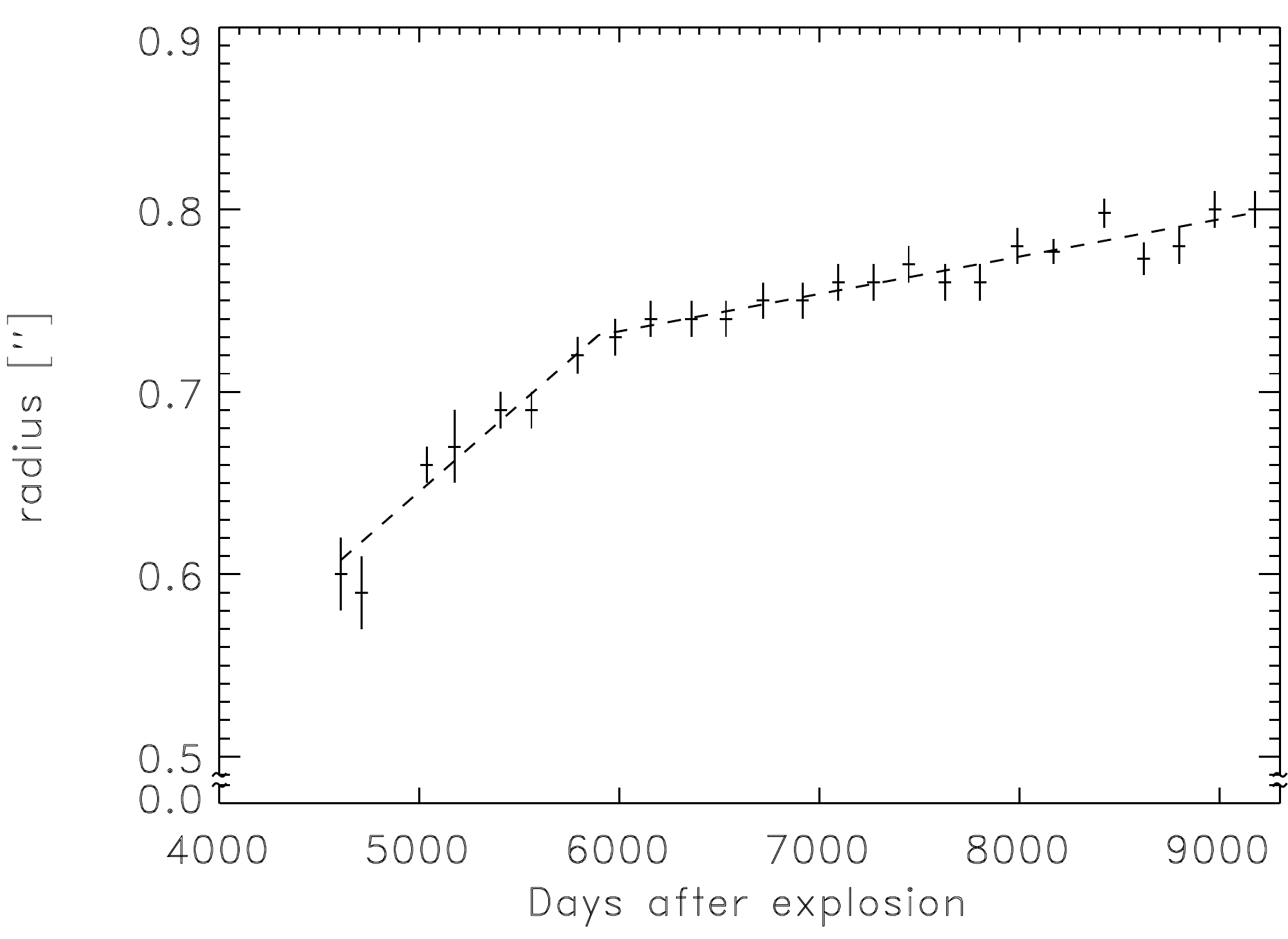}
\caption{Radii of a toroidal model (\S \ref{imaging}) fit to each image in Figure 1. }
\label{radii}
\end{figure}

For the September 2011 and March 2012 observations, we analyze only the longest exposure, to avoid distortions caused by residual alignment errors between exposures. Figure \ref{overview} shows the deconvolved images of all epochs of {\it Chandra} observations to date. To characterize these images, we fit the image with a model that consists of four lobes and a ring \citep{Racusin2009}. Figure \ref{radii} shows the best-fit radii of all the images considered in this study. 

\subsection{Spectra}

Spectra were extracted from each observation within a circular aperture (radius $4\farcs38$) sized to encompass nearly all the light from the target in all epochs, using CIAO version 4.4  \citep{Fruscione06} and CALDB version 4.4.10.  Background spectra were extracted from each observation within an annular region (radius $6\farcs2$ to $12\farcs4$).

\begin{figure}
\epsscale{1.}
\plotone{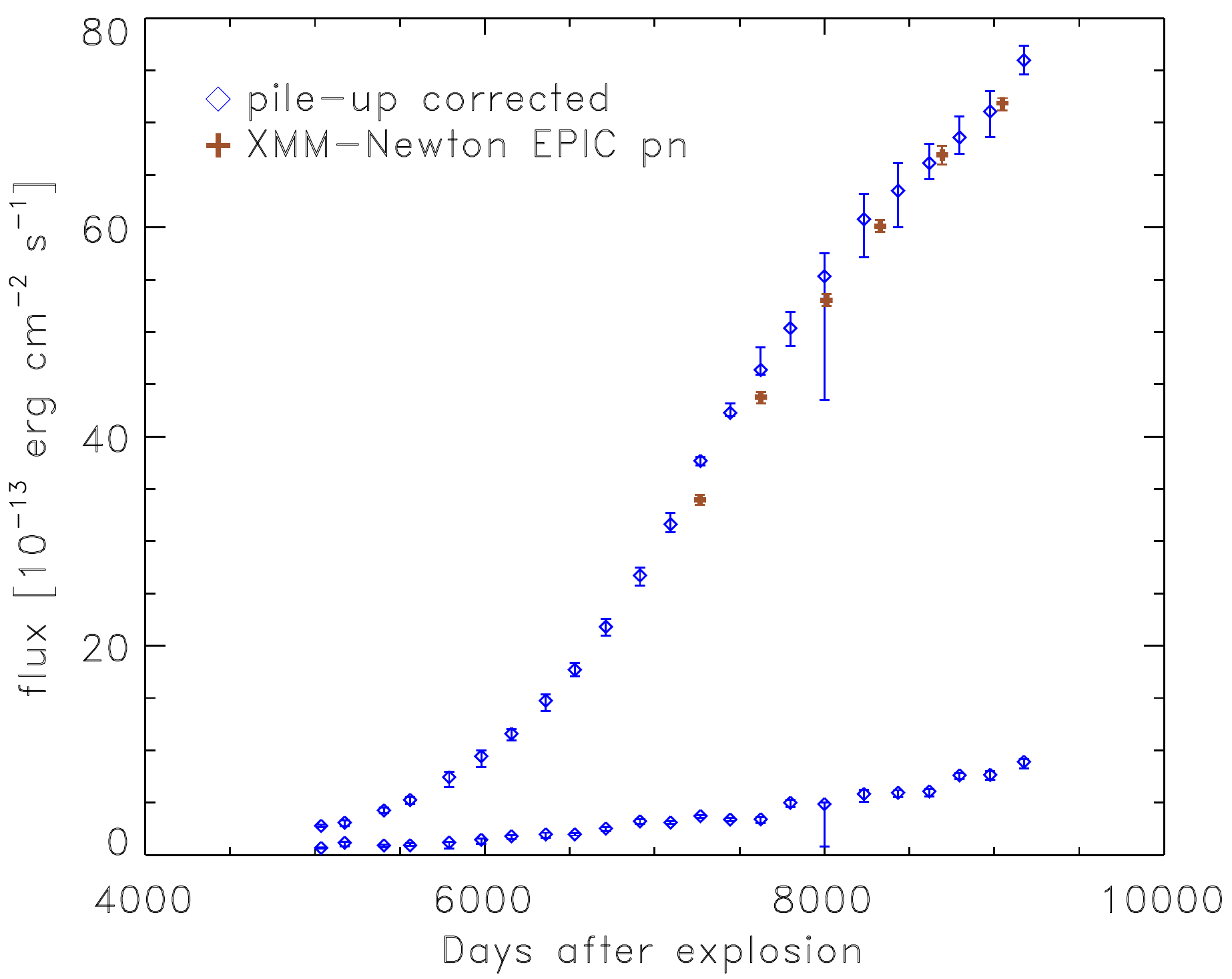}
\caption{{\em Chandra} ACIS light curves for the soft (0.5-2.0 keV, higher curve) and hard X-ray (3.0-8.0 keV, lower curve) bands, with pileup correction (\S2.2). {\it XMM-Newton} EPIC pn fluxes are shown (brown) for comparison.  Uncertainties shown are 1$\sigma$. The {\it XMM-Newton} EPIC pn fluxes are displayed with 3 $\sigma$-errors. }
\label{lightcurve}
\end{figure}

Even though we adjusted the observing modes to minimize pileup, PIMMS suggests that a modest pileup is present in many of the observations.  We first tried to take this into account during the fitting process by using the pileup model described by \cite{Davis2001}. 
This approach adds to the overall model several pileup parameters
which then have to be either assigned appropriate values or
included in the fitting; for details
see the ``Chandra ABC Guide to Pileup"\footnote{http://cxc.harvard.edu/ciao/download/doc/pileup\_abc.pdf,
Section 4.1: Correcting Imaging Observations.}.
As an example, the parameter {\tt nregions} encodes the spatial size of
the source in units of $3\times 3$ pixel cells.  Based on the
ACIS image we can estimate that {\tt nregions} should be in
the range of 2 to 5, but we cannot determine the exact value a priori.
Including {\tt nregions} in the fitting however adds a degeneracy
between {\tt nregions} and the other model parameters such that the fit
and its flux are not well constrained.

An alternative, parameter-free strategy for addressing pileup in the analysis of ACIS data has been described by \citet[][Appendix A]{Broos11}. First, a high-fidelity simulator of the ACIS detector \citep{Townsley02a}, including charge transfer inefficiency \citep{Townsley02b}, is used to infer an input photon spectrum that would cause ACIS to produce the observed spectrum and event rate under the specific configuration of the observation. That photon spectrum is very flexible; no spectral model is assumed. The spatial distribution of the simulated photons incident on ACIS was produced by the MARX package\footnote{\url{http://space.mit.edu/cxc/marx/}}, using the corresponding SN 1987A morphology model described in \S \ref{imaging}.  MARX and the CCD simulation were configured to match the parameters of the observation, such as the target location on the focal plane, frame time, and dead area/time\footnote{
See discussion of {\em FRACEXPO} at \url{http://cxc.harvard.edu/ciao4.4/why/acisdeadarea.html}}. 

 Second, with the input photon spectrum fixed, the simulation is repeated with only a single photon arriving in each frame, eliminating pileup.  The resulting simulated spectrum, which we refer to as ``pileup corrected'', is then fit using standard methods and calibration files. Figure \ref{pileup} shows the original and pileup-corrected spectrum of the observation that suffered worst from pileup (ObsId 4614). We applied this method to all ACIS observations reported here. The flux correction attributed to pileup in each observation is reported in Table \ref{tabel}. This Table shows that the correction for the soft band is more smooth than the correction of the harder band. This probably indicates that our pileup correction method is more stable for the soft band than the hard band.

\subsubsection{Light curve}
\begin{figure}
\epsscale{1.}
\plotone{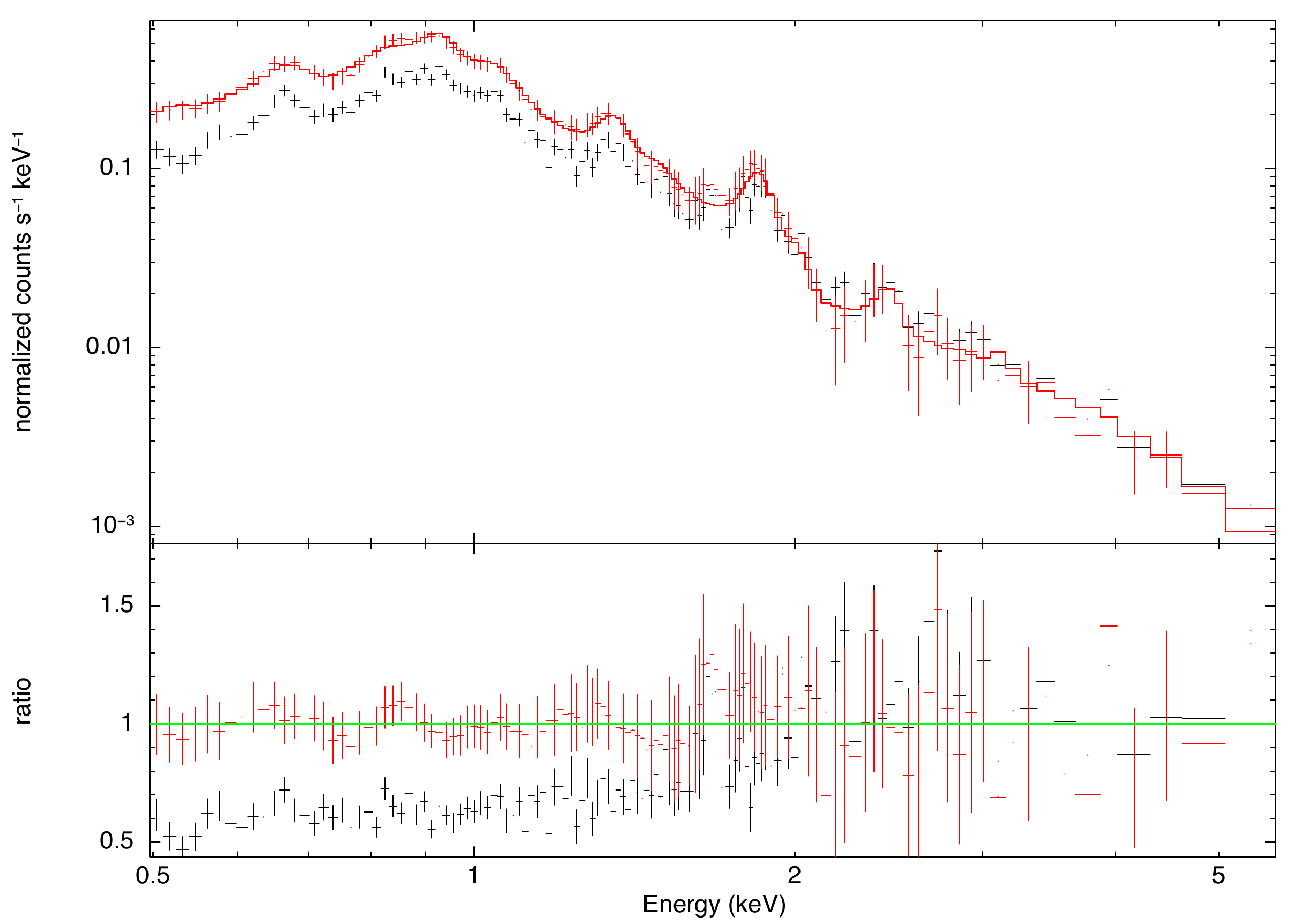}
\caption{Original (black) and pileup-corrected spectrum of ObsId 4614 with best-fit model overplotted. Note that the piledup spectrum is substantially harder than the pileup-corrected spectrum.}
\label{pileup}
\end{figure}

We fitted the resulting pileup-corrected spectra with an absorbed two component spectral model, using the X-ray spectral fitting package XSPEC, version 12.6.0 \citep{xspec}.  The spectral model included a component in collisional equilibrium ionization (\texttt{vequil}) and a non-equilibrium ionization component (\texttt{vpshock}) using \texttt{neivers} 2.0, updated with atomic data for inner shell processes \cite[c.f.][]{Badenes2006}.We fixed the abundances to those found in the high resolution X-ray LETG spectra investigated by \cite{Zhekov2006} and we fixed the absorption to 2.35$\times 10^{21}$ cm$^{-2}$ \citep{Park2006}. Figure \ref{pileup} shows an example of one of the fits. We measured the flux in the 0.5-2.0 and 3.0-8.0 keV energy bands from the best-fit model for all spectra and estimated flux uncertainties for a 68\% confidence level.  
Figure \ref{lightcurve} shows the evolution of the fluxes in the low and high energy bands for the observations corrected for pileup effects. 
For this study, we do not need the high signal-to-noise that would be required for a detailed spectral analysis. Therefore, we decided to only correct the longest observation per epoch for pileup even if there were more observations available in the same epoch. This explains the large error bars of the flux at day 7997, as this epoch was divided into 5 observations. 

For comparison, we overplotted the fluxes of the last six epochs of {\it XMM Newton} EPIC pn fluxes \citep[][]{Maggi12}. Note that this instrument has been shown to be extremely stable \citep[e.g.,][]{Sartore12,Plucinsky2012}.

\subsubsection{Chandra grating observations }
After the previously described procedures, the 0.5-2.0 keV light curve increases more or less linearly from day 6400 on, but there still appears to be a break around day 8200, which does not appear in the {\it XMM-Newton} EPIC pn light curve \citep{Maggi12}. 

Since the pileup corrections (Table \ref{tabel}) after the break are relatively low (7--11\%) compared to some earlier observations, insufficient pileup correction is an unlikely explanation for the break.

\begin{figure}
\epsscale{1.}
\plotone{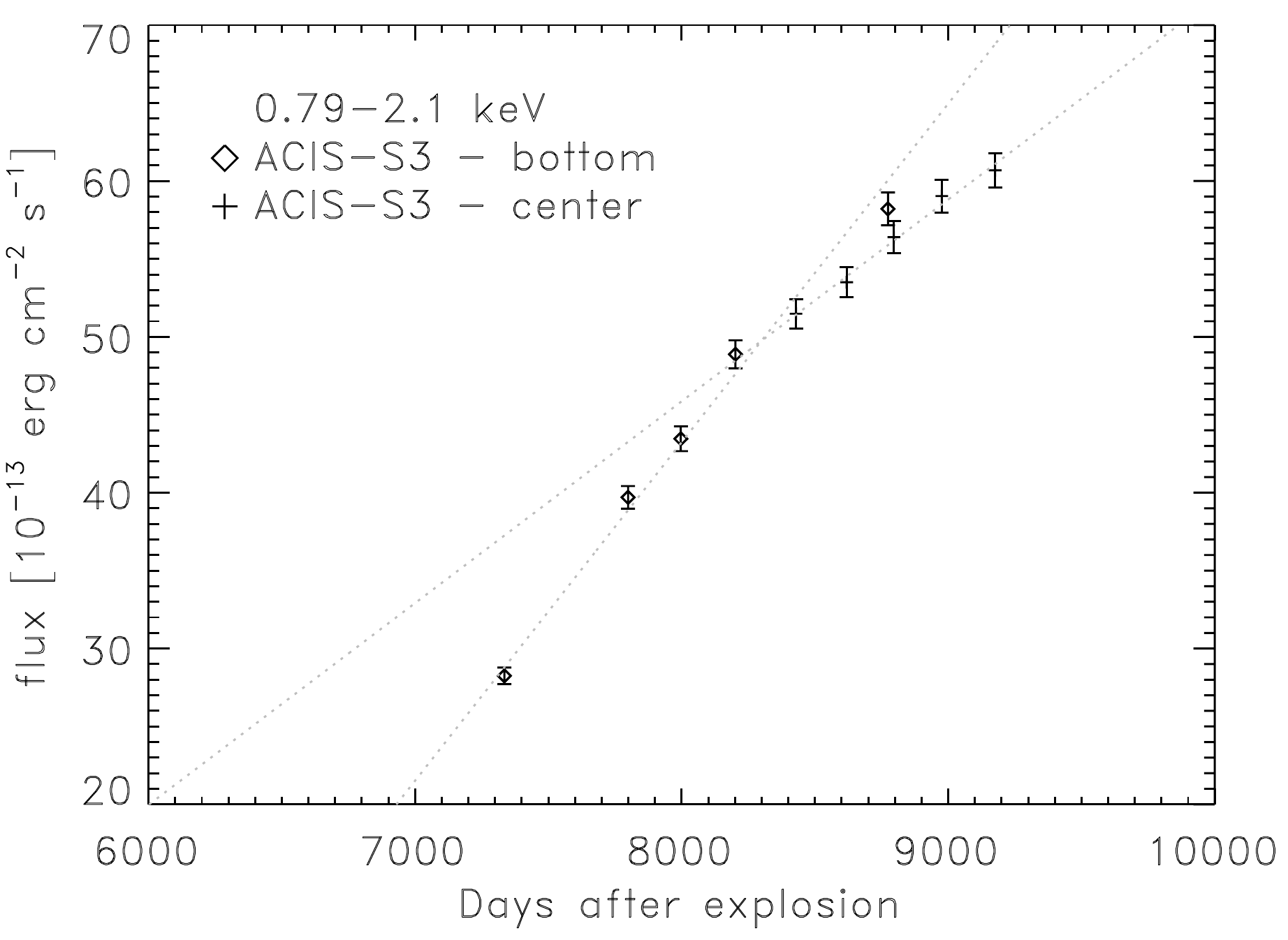}
\caption{{\em Chandra} HETG soft band (0.79-2.1 keV) light curves in the dispersed spectra. Note that around day 8200 there is a suggestion of a break. The lines are fitted to the data points for similar instrument setups individually. Uncertainties shown are 1$\sigma$.}
\label{hetgACIS}
\end{figure}

We can confirm this directly using the data taken with the ACIS-HETG configuration, since the dispersed spectra in each of these observations have no pileup, they provide invaluable cross-checks for our pileup-corrected ACIS fluxes. The dispersed spectra were extracted
using the standard CIAO tools and extraction sizes.  The Medium Energy Grating (MEG) plus and minus data are combined
and the fluxes in the 0.79 to 2.1 keV range are
determined directly from the flux-corrected data,
as was done in \cite{Dewey12}. In addition to the monitoring observations listed in Table \ref{tabel}, we analyzed the deep observations taken in 2007 (day 7335) and 2011 (day 8774). These observations have been described in \citet{Dewey12}. The light curve obtained from the dispersed spectra (Figure \ref{hetgACIS}) shows a break around day 8200 similar to that seen in Figure \ref{lightcurve}. Since this break occurs at the same time that we moved the aimpoint on the CCD, we cannot disentangle any effects caused by that change from those caused by the increased contamination deposition rate, but we can be confident that this break is not caused by the increased count rate and insufficient pileup correction.

\section{Discussion}

\subsection{Soft X-ray emission}

\cite{Haberl2006} predicted a flattening of the 0.5-2.0 keV light curve around day 7000 and predicted the light curve to reach its maximum around day 8000. Also, \citet{Borkowski97} predicted the soft X-ray light curve to reach a maximum around day 9000, but with a flux of at up to a factor of ten {\em above} the current X-ray flux. Indeed, \cite{Park2011} showed that the flux in the 0.5-2.0 keV energy band flattened around day 8200 and argued that this flattening is a result of the transmitted shocks entering a lower-density region. 

The improved calibration of the optical blocking filter now available removes a substantial part of this flattening. However, there still appears to be a break in the {\it Chandra} light curves, both in the piledup data (bare ACIS and zeroth order HETG) and in the pileup-free dispersed data.  Note that all the {\it Chandra} data presented here rely on the same calibration model for the optical blocking filter. The {\it XMM-Newton} EPIC-pn data show no evidence for for a sudden break around day 8200.
 Given that this break is visible in both dispersed and zeroth order, but not in the {\it XMM-Newton} EPIC pn data, we conclude that this is not likely an effect of an insufficient pileup correction, but rather is caused by a miscalibration of the increase in contamination on the optical blocking filters, combined with a change in instrument setup. Note that this contamination has a wavelength dependent effect on the spectrum and hence has influence on the best-fit spectral parameters. Also, this contamination has less of an effect on the 3.0-8.0 keV flux.  
 
 \subsection{Imaging}
Figure \ref{overview} shows the evolution of SN 1987A's morphology. To interpret these images, one has to keep in mind that some of the structure in the images might be an artifact of the deconvolution method we used to correct for the extension of the point spread function of {\it Chandra}. Moreover, the point spread function of {\it Chandra} is not entirely symmetric, as since December 2000, there appeared a hook-shaped extension to the point spread function\footnote{see \url{http://cxc.harvard.edu/ciao4.4/caveats/psf\_artifact.html}}, which is not accounted for in MARX 4.4. This hook contains about 6\% of the total flux and will probably show up after the deconvolution as a small extending structure to the remnant.  

The individual spots on the west side of the remnant appear to have merged in the last three epochs. 
One question that comes to mind when looking at Figure \ref{radii} is ``how does this relate to the inner-ring structure as observed by \cite{Jakobsen91}?''. \cite{Ng2009} compared radii of X-ray and radio images of SN 1987A and found comparable sizes, with radii larger than the radius of the equatorial ring in 2008. However, \cite{Dewey12} showed a strong break in the evolution of the soft X-ray light curve after the shocks have overcome the equatorial ring, and we do not yet see such evolution in the 0.5-2.0 keV light curve presented in Figure \ref{lightcurve}. 

To estimate how far the shocks progressed through the equatorial ring, we need to know when the interaction started. \cite{Sugerman2002} carried out a study on multiple epochs of Hubble Space Telescope (HST) images, ranging from 1994 February and 2000 February. The first spot lit up by the blast wave is located in the northeast quadrant of the remnant and was detected in a HST observations taken 2933 days after the explosion. Remarkably, this spot is located 0.06$''$ inside of the main ring as visible in optical  \citep{Sugerman2002}. The second spot was only found in a HST observation taken 1350 days later and by day 4999, 12 spots were visible in the HST observations \citep{Sugerman2002}. A lower limit to the width of the ring, as visible in the HST image, is of the order of $0\farcs12$ \cite[$\sim 10^{12}$ km,][]{Plait1995}.  
Fitting a broken linear function to the radii of Figure \ref{radii} results in a velocity of $8500^{+2300}_{-1900}$ km~s$^{-1}$ until day $5900^{+280}_{-230}$, after which the velocity is $1820^{+350}_{-380}$ km~s$^{-1}$. Assuming that the shock wave reached the ring at day 4999 when 12 spots were lit up (hereby ignoring the spots that were hit first, as those were protruding inwards), the radius has increased by $(1.17\pm 0.06) \times 10^{12}$ km \citep[$\sim$120\% of the width reported by][]{Plait1995} since the interaction started. We used a Monte Carlo simulation to estimate the uncertainties on this value.
Note that this distance relies on the assumption that the 12 spots reported by \cite{Sugerman2002} are part of the ring reported by \cite{Plait1995}.

\cite{Dwek2010} argued, based on the evolution of the infrared emission, that the forward shock only fully started to interact with the equatorial ring at day $\sim$5600. This appears consistent with the increase in soft X-ray flux around day 6000 \citep{Park05}. The latter was considered by \cite{Dewey12} to be consistent with the lighting up of the optical spots $\sim$ 1000 days earlier, as their models show some delay between the shock impact of dense material and the onset of X-ray emission. Nevertheless, if we take day 5600 as conservative assumption for the start of the explosion to interact with the ring, we estimate the increase of the radius of the X-ray emission since then to be $7.36\pm 0.05\times10^{11}$ km \citep[$\sim$74\% of the width reported by][]{Plait1995}. 

It is not straightforward to interpret the increase of the radius of the X-ray emission in terms of the 3 components as described in the introduction. The velocity is too high for the component with low ($\sim$ 0.3-0.5 keV) temperature \citep{Zhekov09}. \cite{Dewey12} reported the contribution of a high velocity component to be $\sim 20$ \% in the HETG spectra, but its contribution to the broadband flux might be higher. Note that they measured the spectral lines to be much broader than 1700 km s$^{-1}$. Therefore, the measured increase of the radius is likely a combination of the increase in radius of all three components.  The radius of the X-ray emission is identified in \cite{Dewey12} as following the
progress of the forward shock/contact discontinuity as it moves through the uniform-with-clumps equatorial ring.
Figure 9 (top) of \cite{Dewey12} shows the forward shock/contact discontinuity locations along with the \citet{Racusin2009} radii and there is reasonable agreement.  The speed of the forward shock there is of order 1700 km~s$^{-1}$.

Current and future HST observations will reveal how far the shock has progressed, and how many of the optical spots have merged by now. A combined study with optical, X-ray and infrared data will reveal a detailed picture of the structure of this equatorial ring, leading to better constraints on the last stages in the life of the progenitor of SN 1987A.

\section{Conclusions}
We report our results for the three newest epochs of {\it Chandra} data on SN 1987A. Together with our reanalysis of older epochs, we paint a picture of the evolution of SN 1987A as witnessed by {\it Chandra} in the past 10 years. Based on this analysis, we reach the following conclusions:
\begin{itemize}

\item[-] The increase of the radius of the ring visible in X-rays since the explosion first interacted with the preexisting equatorial ring is 74-120\% of the lower limit to the thickness of this preexisting equatorial ring as measured by \cite{Plait1995}.

\item[-] Given that the shocks have now traveled through a substantial amount of the preexisting equatorial ring, a multiwavelength study including high spatial resolution optical images is necessary to investigate how far the shock has traveled through the ring.

\item[-] The sudden break in the 0.5-2.0 keV light curve around day 8200 reported by \cite{Park2011} was probably an instrumental effect. 
The absence of a break in the light curve likely also indicates that the shocks traveling through the dense parts of the ring have not fully overcome the extent of the ring yet.

\end{itemize}

\acknowledgements{We thank Frank Haberl and Pierre Maggi for providing us with the XMM-Newton EPIC pn fluxes in advance of publication. We thank Herman Marshall, Paul Plucinsky, Konstatin Getman, Bettina Posselt, Zachary Prieskorn, Jonathan Gelbord and Binbin Zhang for discussions about statistics, calibration and pileup corrections. This work is supported by the ACIS Instrument Team contract SV4-74018 (PI:  G.\ Garmire), issued by the {\em Chandra} X-ray Center, which is operated by the Smithsonian Astrophysical Observatory for and on behalf of NASA under contract NAS8-03060. E.A.H. and D.N.B. are supported by SAO grants GO1-12070X and GO2-13064X.}


\begin{thebibliography}{39}
\expandafter\ifx\csname natexlab\endcsname\relax\def\natexlab#1{#1}\fi

\bibitem[{{Arnaud}(1996)}]{xspec}
{Arnaud}, K.~A. 1996, in Astronomical Society of the Pacific Conference Series,
  Vol. 101, Astronomical Data Analysis Software and Systems V, ed.
  {G.~H.~Jacoby \& J.~Barnes}, 17

\bibitem[{{Badenes} {et~al.}(2006){Badenes}, {Borkowski}, {Hughes}, {Hwang}, \&
  {Bravo}}]{Badenes2006}
{Badenes}, C., {Borkowski}, K.~J., {Hughes}, J.~P., {Hwang}, U., \& {Bravo}, E.
  2006, \apj, 645, 1373

\bibitem[{{Blondin} \& {Lundqvist}(1993)}]{Blondin93}
{Blondin}, J.~M. \& {Lundqvist}, P. 1993, \apj, 405, 337

\bibitem[{{Borkowski} {et~al.}(1997){Borkowski}, {Blondin}, \&
  {McCray}}]{Borkowski97}
{Borkowski}, K.~J., {Blondin}, J.~M., \& {McCray}, R. 1997, \apj, 477, 281

\bibitem[{{Broos} {et~al.}(2011){Broos}, {Townsley}, {Feigelson}, {Getman},
  {Garmire}, {Preibisch}, {Smith}, {Babler}, {Hodgkin}, {Indebetouw}, {Irwin},
  {King}, {Lewis}, {Majewski}, {McCaughrean}, {Meade}, \&
  {Zinnecker}}]{Broos11}
{Broos}, P.~S., {Townsley}, L.~K., {Feigelson}, E.~D., {Getman}, K.~V.,
  {Garmire}, G.~P., {Preibisch}, T., {Smith}, N., {Babler}, B.~L., {Hodgkin},
  S., {Indebetouw}, R., {Irwin}, M., {King}, R.~R., {Lewis}, J., {Majewski},
  S.~R., {McCaughrean}, M.~J., {Meade}, M.~R., \& {Zinnecker}, H. 2011, \apjs,
  194, 2

\bibitem[{{Burrows} {et~al.}(2000){Burrows}, {Michael}, {Hwang}, {McCray},
  {Chevalier}, {Petre}, {Garmire}, {Holt}, \& {Nousek}}]{Burrows}
{Burrows}, D.~N., {Michael}, E., {Hwang}, U., {McCray}, R., {Chevalier}, R.~A.,
  {Petre}, R., {Garmire}, G.~P., {Holt}, S.~S., \& {Nousek}, J.~A. 2000, \apjl,
  543, L149

\bibitem[{{Chita} {et~al.}(2008){Chita}, {Langer}, {van Marle},
  {Garc{\'{\i}}a-Segura}, \& {Heger}}]{Chita2008}
{Chita}, S.~M., {Langer}, N., {van Marle}, A.~J., {Garc{\'{\i}}a-Segura}, G.,
  \& {Heger}, A. 2008, \aap, 488, L37

\bibitem[{{Crotts} {et~al.}(1989){Crotts}, {Kunkel}, \& {McCarthy}}]{Crotts89}
{Crotts}, A.~P.~S., {Kunkel}, W.~E., \& {McCarthy}, P.~J. 1989, \apjl, 347, L61

\bibitem[{{Davis}(2001)}]{Davis2001}
{Davis}, J.~E. 2001, \apj, 562, 575

\bibitem[{{Dewey} {et~al.}(2012){Dewey}, {Dwarkadas}, {Haberl}, {Sturm}, \&
  {Canizares}}]{Dewey12}
{Dewey}, D., {Dwarkadas}, V.~V., {Haberl}, F., {Sturm}, R., \& {Canizares},
  C.~R. 2012, \apj, 752, 103

\bibitem[{{Dwek} {et~al.}(2010){Dwek}, {Arendt}, {Bouchet}, {Burrows},
  {Challis}, {Danziger}, {De Buizer}, {Gehrz}, {Park}, {Polomski}, {Slavin}, \&
  {Woodward}}]{Dwek2010}
{Dwek}, E., {Arendt}, R.~G., {Bouchet}, P., {Burrows}, D.~N., {Challis}, P.,
  {Danziger}, I.~J., {De Buizer}, J.~M., {Gehrz}, R.~D., {Park}, S.,
  {Polomski}, E.~F., {Slavin}, J.~D., \& {Woodward}, C.~E. 2010, \apj, 722, 425

\bibitem[{{Fruscione} {et~al.}(2006){Fruscione}, {McDowell}, {Allen},
  {Brickhouse}, {Burke}, {Davis}, {Durham}, {Elvis}, {Galle}, {Harris},
  {Huenemoerder}, {Houck}, {Ishibashi}, {Karovska}, {Nicastro}, {Noble},
  {Nowak}, {Primini}, {Siemiginowska}, {Smith}, \& {Wise}}]{Fruscione06}
{Fruscione}, A., {McDowell}, J.~C., {Allen}, G.~E., {Brickhouse}, N.~S.,
  {Burke}, D.~J., {Davis}, J.~E., {Durham}, N., {Elvis}, M., {Galle}, E.~C.,
  {Harris}, D.~E., {Huenemoerder}, D.~P., {Houck}, J.~C., {Ishibashi}, B.,
  {Karovska}, M., {Nicastro}, F., {Noble}, M.~S., {Nowak}, M.~A., {Primini},
  F.~A., {Siemiginowska}, A., {Smith}, R.~K., \& {Wise}, M. 2006, in Society of
  Photo-Optical Instrumentation Engineers (SPIE) Conference Series, Vol. 6270,
  Society of Photo-Optical Instrumentation Engineers (SPIE) Conference Series

\bibitem[{{Haberl} {et~al.}(2006){Haberl}, {Geppert}, {Aschenbach}, \&
  {Hasinger}}]{Haberl2006}
{Haberl}, F., {Geppert}, U., {Aschenbach}, B., \& {Hasinger}, G. 2006, \aap,
  460, 811

\bibitem[{{Jakobsen} {et~al.}(1991){Jakobsen}, {Albrecht}, {Barbieri},
  {Blades}, {Boksenberg}, {Crane}, {Deharveng}, {Disney}, {Kamperman}, {King},
  {Macchetto}, {Mackay}, {Paresce}, {Weigelt}, {Baxter}, {Greenfield},
  {Jedrzejewski}, {Nota}, {Sparks}, {Kirshner}, \& {Panagia}}]{Jakobsen91}
{Jakobsen}, P., {Albrecht}, R., {Barbieri}, C., {Blades}, J.~C., {Boksenberg},
  A., {Crane}, P., {Deharveng}, J.~M., {Disney}, M.~J., {Kamperman}, T.~M.,
  {King}, I.~R., {Macchetto}, F., {Mackay}, C.~D., {Paresce}, F., {Weigelt},
  G., {Baxter}, D., {Greenfield}, P., {Jedrzejewski}, R., {Nota}, A., {Sparks},
  W.~B., {Kirshner}, R.~P., \& {Panagia}, N. 1991, \apjl, 369, L63

\bibitem[{{Lucy}(1974)}]{Lucy}
{Lucy}, L.~B. 1974, \aj, 79, 745

\bibitem[{{Luo} \& {McCray}(1991)}]{Luo91}
{Luo}, D. \& {McCray}, R. 1991, \apj, 379, 659

\bibitem[{{Maggi} {et~al.}(2012){Maggi}, {Haberl}, {Sturm}, \&
  {Dewey}}]{Maggi12}
{Maggi}, P., {Haberl}, F., {Sturm}, R., \& {Dewey}, D. 2012, \aap, accepted.
  ArXiv: 1211.1220

\bibitem[{{Mattila} {et~al.}(2010){Mattila}, {Lundqvist}, {Gr{\"o}ningsson},
  {Meikle}, {Stathakis}, {Fransson}, \& {Cannon}}]{Matilla10}
{Mattila}, S., {Lundqvist}, P., {Gr{\"o}ningsson}, P., {Meikle}, P.,
  {Stathakis}, R., {Fransson}, C., \& {Cannon}, R. 2010, \apj, 717, 1140

\bibitem[{{Mori} {et~al.}(2001){Mori}, {Tsunemi}, {Miyata}, {Baluta},
  {Burrows}, {Garmire}, \& {Chartas}}]{Mori2001}
{Mori}, K., {Tsunemi}, H., {Miyata}, E., {Baluta}, C.~J., {Burrows}, D.~N.,
  {Garmire}, G.~P., \& {Chartas}, G. 2001, in Astronomical Society of the
  Pacific Conference Series, Vol. 251, New Century of X-ray Astronomy, ed.
  {H.~Inoue \& H.~Kunieda}, 576

\bibitem[{{Morris} \& {Podsiadlowski}(2007)}]{Morris07}
{Morris}, T. \& {Podsiadlowski}, P. 2007, Science, 315, 1103

\bibitem[{{Morris} \& {Podsiadlowski}(2009)}]{Morris2009}
---. 2009, \mnras, 399, 515

\bibitem[{{Ng} {et~al.}(2009){Ng}, {Gaensler}, {Murray}, {Slane}, {Park},
  {Staveley-Smith}, {Manchester}, \& {Burrows}}]{Ng2009}
{Ng}, C.-Y., {Gaensler}, B.~M., {Murray}, S.~S., {Slane}, P.~O., {Park}, S.,
  {Staveley-Smith}, L., {Manchester}, R.~N., \& {Burrows}, D.~N. 2009, \apjl,
  706, L100

\bibitem[{{Park} {et~al.}(2006){Park}, {Zhekov}, {Burrows}, {Garmire},
  {Racusin}, \& {McCray}}]{Park2006}
{Park}, S., {Zhekov}, S.~A., {Burrows}, D.~N., {Garmire}, G.~P., {Racusin},
  J.~L., \& {McCray}, R. 2006, \apj, 646, 1001

\bibitem[{{Park} {et~al.}(2005){Park}, {Zhekov}, {Burrows}, \&
  {McCray}}]{Park05}
{Park}, S., {Zhekov}, S.~A., {Burrows}, D.~N., \& {McCray}, R. 2005, \apjl,
  634, L73

\bibitem[{{Park} {et~al.}(2011){Park}, {Zhekov}, {Burrows}, {Racusin}, {Dewey},
  \& {McCray}}]{Park2011}
{Park}, S., {Zhekov}, S.~A., {Burrows}, D.~N., {Racusin}, J.~L., {Dewey}, D.,
  \& {McCray}, R. 2011, \apjl, 733, L35

\bibitem[{{Plait} {et~al.}(1995){Plait}, {Lundqvist}, {Chevalier}, \&
  {Kirshner}}]{Plait1995}
{Plait}, P.~C., {Lundqvist}, P., {Chevalier}, R.~A., \& {Kirshner}, R.~P. 1995,
  \apj, 439, 730

\bibitem[{{Plucinsky} {et~al.}(2012)}]{Plucinsky2012}
{Plucinsky}, P.~P. {et~al.} 2012, in Society of Photo-Optical Instrumentation
  Engineers (SPIE) Conference Series, Vol. 8443, Society of Photo-Optical
  Instrumentation Engineers (SPIE) Conference Series

\bibitem[{{Racusin} {et~al.}(2009){Racusin}, {Park}, {Zhekov}, {Burrows},
  {Garmire}, \& {McCray}}]{Racusin2009}
{Racusin}, J.~L., {Park}, S., {Zhekov}, S., {Burrows}, D.~N., {Garmire}, G.~P.,
  \& {McCray}, R. 2009, \apj, 703, 1752

\bibitem[{{Richardson}(1972)}]{Richardson}
{Richardson}, W.~H. 1972, Journal of the Optical Society of America
  (1917-1983), 62, 55

\bibitem[{{Sartore} {et~al.}(2012){Sartore}, {Tiengo}, {Mereghetti}, {De Luca},
  {Turolla}, \& {Haberl}}]{Sartore12}
{Sartore}, N., {Tiengo}, A., {Mereghetti}, S., {De Luca}, A., {Turolla}, R., \&
  {Haberl}, F. 2012, \aap, 541, A66

\bibitem[{{Sugerman} {et~al.}(2002){Sugerman}, {Lawrence}, {Crotts}, {Bouchet},
  \& {Heathcote}}]{Sugerman2002}
{Sugerman}, B.~E.~K., {Lawrence}, S.~S., {Crotts}, A.~P.~S., {Bouchet}, P., \&
  {Heathcote}, S.~R. 2002, \apj, 572, 209

\bibitem[{{Townsley} {et~al.}(2002{\natexlab{a}}){Townsley}, {Broos},
  {Chartas}, {Moskalenko}, {Nousek}, \& {Pavlov}}]{Townsley02a}
{Townsley}, L.~K., {Broos}, P.~S., {Chartas}, G., {Moskalenko}, E., {Nousek},
  J.~A., \& {Pavlov}, G.~G. 2002{\natexlab{a}}, Nuclear Instruments and Methods
  in Physics Research A, 486, 716

\bibitem[{{Townsley} {et~al.}(2002{\natexlab{b}}){Townsley}, {Broos}, {Nousek},
  \& {Garmire}}]{Townsley02b}
{Townsley}, L.~K., {Broos}, P.~S., {Nousek}, J.~A., \& {Garmire}, G.~P.
  2002{\natexlab{b}}, Nuclear Instruments and Methods in Physics Research A,
  486, 751

\bibitem[{{Townsley} {et~al.}(2003){Townsley}, {Feigelson}, {Montmerle},
  {Broos}, {Chu}, \& {Garmire}}]{Townsley03}
{Townsley}, L.~K., {Feigelson}, E.~D., {Montmerle}, T., {Broos}, P.~S., {Chu},
  Y.-H., \& {Garmire}, G.~P. 2003, \apj, 593, 874

\bibitem[{{Wang} \& {Mazzali}(1992)}]{Wang92}
{Wang}, L. \& {Mazzali}, P.~A. 1992, \nat, 355, 58

\bibitem[{{Zhekov} {et~al.}(2005){Zhekov}, {McCray}, {Borkowski}, {Burrows}, \&
  {Park}}]{Zhekov05}
{Zhekov}, S.~A., {McCray}, R., {Borkowski}, K.~J., {Burrows}, D.~N., \& {Park},
  S. 2005, \apjl, 628, L127

\bibitem[{{Zhekov} {et~al.}(2006){Zhekov}, {McCray}, {Borkowski}, {Burrows}, \&
  {Park}}]{Zhekov2006}
---. 2006, \apj, 645, 293

\bibitem[{{Zhekov} {et~al.}(2009){Zhekov}, {McCray}, {Dewey}, {Canizares},
  {Borkowski}, {Burrows}, \& {Park}}]{Zhekov09}
{Zhekov}, S.~A., {McCray}, R., {Dewey}, D., {Canizares}, C.~R., {Borkowski},
  K.~J., {Burrows}, D.~N., \& {Park}, S. 2009, \apj, 692, 1190

\bibitem[{{Zhekov} {et~al.}(2010){Zhekov}, {Park}, {McCray}, {Racusin}, \&
  {Burrows}}]{Zhekov10}
{Zhekov}, S.~A., {Park}, S., {McCray}, R., {Racusin}, J.~L., \& {Burrows},
  D.~N. 2010, \mnras, 407, 1157

\end{thebibliography}
\end{document}